# Why is Seasonal Climate Predictable Beyond the Limit of Deterministic Predictability set by Chaos?

Vladimir García-Morales[1,2], Devabrat Sharma[2,3], Shruti Tandon[2,3], B. N. Goswami[4], and R. I. Sujith[2,3]

[1]Departament de Física de la Terra i Termodinàmica, Universitat de València, E-46100 Burjassot, Spain
[2]Department of Aerospace Engineering, Indian Institute of Technology Madras, Chennai-600036, India
[3]Centre of Excellence for Studying Critical Transitions in Complex Systems, Indian Institute of Technology Madras, Chennai-600036, India
[4]ST Radar Centre, Gauhati University, Guwahati-781014, India

*Corresponding author. Email: sujith@iitm.ac.in

## Abstract

The Earth's climate is an ensemble of interacting, spatially extended oscillatory media ('climate systems') whose slow-changing averages coexist with chaotic, high-frequency weather fluctuations in quasi-equilibrium. The limit of deterministic predictability (LDP) for any climate system is determined by its fastest-growing errors. However, recent findings show that the Indian Summer Monsoon Rainfall (ISMR) can be predicted up to 18 months in advance—far beyond its LDP. Using a model of two interacting oscillatory media, we show that this extended predictability arises from lag synchronization between ISMR and its predictor, the Global El Niño–Southern Oscillation (G-ENSO), to which it is strongly coupled. We introduce complex order parameters representing the internal dynamics of the two climate systems. Their spatiotemporal evolution is governed by coupled Complex Ginzburg–Landau Equations, producing aperiodic yet strongly correlated time series at long lead times. Our findings have far-reaching consequences in advancing seasonal prediction across climate systems.

## Introduction

The Earth's climate cannot exist without the weather disturbances that transport the excess heat from the tropics to the polar regions and maintain a climate in a dynamic equilibrium (Peixoto & Oort, 1984; Wallace et al., 2023). Thus, the regional weather (all sub-seasonal fluctuations) and the climate (mean state of the region, typically over a season) continuously interact with each other in time. The climate has a dominant annual cycle with the primary external driver being the solar energy. However, the Asian monsoon, the El Niño-Southern Oscillation (ENSO), the Pacific Decadal Oscillation (PDO) or the North Atlantic Oscillation (NAO) have also inter-annual and decadal variability not related to solar irradiance variations. The latter arises from feedback between the systems of the earth's climate. A seminal example of internal climate variability is the ENSO arising from an interaction between the atmosphere and the ocean over the Pacific (Bjerknes, 1969; Philander, 1983). Different regional climates have different spatio-temporal variability and interact with each other bidirectionally. For example, while the ENSO (Philander, 1983; Timmermann et al., 2018; Wang et al., 2017) is considered the primary driver of the Indian summer monsoon rainfall (ISMR) variability (Rasmusson & Carpenter, 1983; Shukla & Paolino, 1983; Walker, 1924; Webster et al., 1998), there is tangible evidence that ISMR may also influence the evolution of the ENSO cycle (Kirtman & Shukla, 2000). Thus, the predictability horizon or the length of lead time up to which useful predictions could be made would depend on the nonlinear interaction of the climate systems.

While weather is a deterministic chaotic system (Lorenz, 1963, 1969) in the high frequency domain with dominant period in days, the regional climate systems are deterministic chaotic systems in the low frequency domain with a dominant period of a few years. The limit on deterministic predictability (LDP) of climate is determined by the growth of errors in the climate system just like for weather (Keane et al., 2025; Lorenz, 1969) although it remains poorly explored. It is solely influenced by the dynamics of their internal degrees of freedom (Philander, 1983; Wang et al., 2017) and was estimated for the ENSO (Andrew Moore & Kleeman, 1996; Battisti et al., 1989; Goswami & Shukla, 1991; Griffies & Bryan, 1997; Latif, 1998; M. Benno Blumenthal, 1991; Webster, 1995).

Recently, by using a novel predictor discovery algorithm (Sharma et al., 2022) pointed out that the Global ENSO of the whole inter-tropical strip (G-ENSO) may provide the basis for long-lead predictability for ISMR up to two years in advance. They showed that a predictor of ISMR based on G-ENSO at 18-month lead (Dp) is correlated with ISMR (Figure 1b) and demonstrated that potential predictability could be realized by using an AI/ML model trained on CMIP6 model simulations and fine-tuned with observations (Sharma et al., 2025). The methodology is promising for correlating G-ENSO to other monsoon or extra-tropical systems. In addition to the interannual variability associated with G-ENSO, it may be noted that ISMR has a multi-decadal variability associated with the North Atlantic Oscillation (NAO) (Borah et al., 2020; Goswami et al., 2006; Rajesh & Goswami, 2020). The length of the time series to represent the multi-decadal variability reliably should be carefully addressed. While for the ISMR we have about 150 years of rainfall data, for the predictor Dp, sufficient length of D20 analysis is available only from an older reanalysis (SODA) (Carton & Giese, 2008; Giese & Ray, 2011).

*Is there predictability of climate beyond the LDP?* This question has become relevant after the recent studies (Sharma et al., 2022, 2025) mentioned above. In this letter, *we affirmatively answer this question* showing how incoherent patterns governing the spatiotemporal behavior of the internal degrees of freedom of climate systems can synchronize to a lead time that is well beyond their LDP. We formulate a model consisting of two coupled Complex Ginzburg-Landau Equations (CGLEs) that govern the evolution of model theoretical descriptors of the ISMR and its predictor. The model yields two aperiodic, yet strongly correlated time series to 18 months lead, far beyond the LDP of the ISMR.

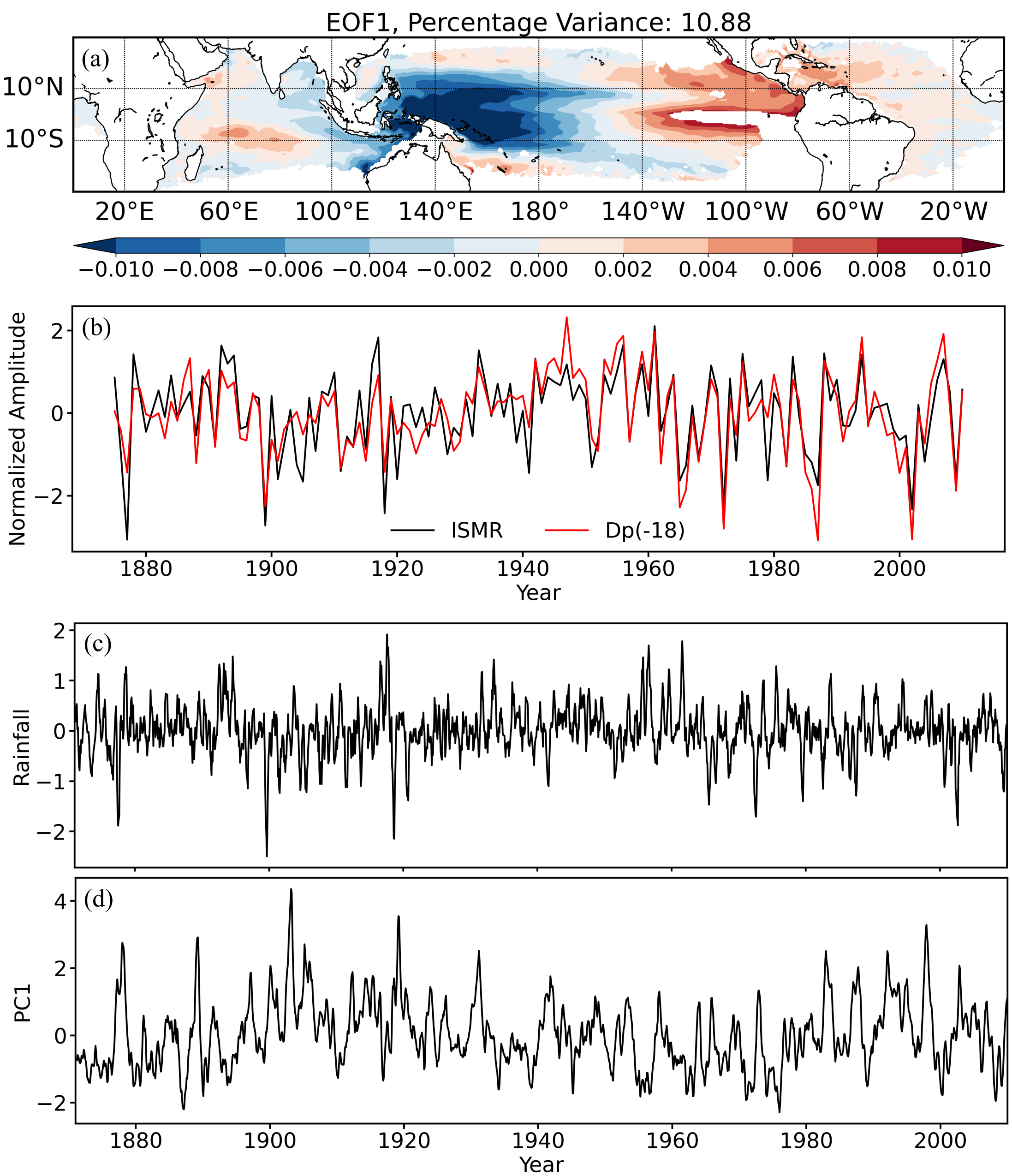

**Figure 1: Global-ENSO predictor of ISMR. (a)** The spatial pattern of the first empirical orthogonal function (EOF1) of monthly mean D20 anomalies from SODA reanalysis for the period 1871 to 2010. **(b)** Annual variation of ISMR and Dp(-18) with strong correlation $r = 0.84$. **(c)** 5-month running mean monthly rainfall anomalies for all India from Indian Institute of Tropical Meteorology (IITM) (Parthasarathy et al., 1994) dataset from 1871-2010. **(d)** Time series of the first Principal Component (PC1) associated with EOF1of D20 anomalies from 1871-2010.

Synchronization of interacting nonlinear oscillators is a well-established concept (Duane, 2015; Kocarev & Parlitz, 1996; Pecora & Carroll, 1990; Pikovsky et al., 2001; Rosenblum et al., 1996, 1997; Zhou, 2006) There also exist studies on synchronization of low-dimensional systems relevant to weather and climate prediction (Duane et al., 2006; Duane, 1997; Gregory S. Duane et al., 1999; Duane & Tribbia, 2001; Duane & Tribbia, 2004; Yang et al., 2006). However, synchronization of incoherent patterns in *ensembles* of *spatially extended oscillatory media* (i.e. high-dimensional systems) remains to be better explored.

Perfect synchrony of incoherent (turbulent) spatial patterns was reported by Zhou (2006). Here we present a more general possibility: *lag synchronization,* with specific time intervals between incoherent spatial patterns. Although this behavior looks superficially similar to so-called *anticipated synchronization* reported in (Ciszak et al., 2015), there is no time delay introduced *ad hoc* in our model (it emerges dynamically in a natural way) and our coupling is bidirectional. When lag synchronization arises in the ensemble of spatially extended oscillatory media, this provides a basis for long-lead predictability of climate systems even in the presence of incoherence. Neither the existence of lag synchronization *in ensembles* of climate systems nor its link to long-lead predictability of the separate systems has ever been addressed.

The CGLE (Aranson & Kramer, 2002; Cross & Hohenberg, 1993; García-Morales & Krischer, 2012) is the model building block of each spatially extended oscillatory medium within the ensemble. In fluid mechanical systems, the CGLE describes the universal behavior of the transition from noise to regular oscillations (García-Morales et al., 2024). In the context of weather and climate systems, the CGLE was used to develop a machine-learning-based prediction framework (Jiang et al., 2022) and to model cloud pattern formation (Monroy & Naumis, 2021).

## 2 Data and Methods

### 2.1 Data

The ISMR is defined as the total rainfall accumulated over the Indian landmass during June-September (JJAS). The ISMR dataset is obtained from a fixed network of 306 rain stations distributed across the Indian landmass, excluding hilly regions such as Jammu and Kashmir, Himachal Pradesh, West Uttar Pradesh Hills, and Arunachal Pradesh (Parthasarathy et al., 1994). The observations from this network are available at a monthly temporal resolution for the period 1871-2016. To construct the G-ENSO index, the depth of the 20°C isotherm (D20) over the global tropics (0°-360°E, 30°S-30°N) is obtained from Simple Ocean Data Assimilation (SODA) version 2.2.4 (Carton & Giese, 2008; Giese & Ray, 2011). Several versions of the SODA datasets exist, differing in their data assimilation configuration (Carton et al., 2018; Carton & Giese, 2008). The output variables from SODA version 2.2.4, include ocean temperature, salinity, horizontal and vertical ocean velocity, sea level, and wind stress. These variables are available as monthly averages on a global grid with a spatial resolution of 0.5° x 0.5°, with 40 vertical levels, covering the period 1871-2010. Monthly anomalies are obtained by removing the climatological mean annual cycle. For gridded dataset, this procedure is applied independently at each grid point. Further, linear trends are removed from all monthly anomalies.

**2.2 Predictor discovery Algorithm for G-ENSO Predictors**

The influence of ENSO on ISMR at a given lead time depends on the global tropical sea surface temperature (SST) configuration at that lead. Consequently, the regional Niño 3.4 index alone is insufficient to represent the full impact of ENSO on ISMR. The correlation between Niño 3.4 and ISMR captures only a partial teleconnection and may become misleading at longer lead times. Recognizing these limitations and the role of the global tropical oceans (0–360°E, 30°S–30°N), (Sharma et al., 2022) introduced the G-ENSO for studying teleconnections with ISMR. The G-ENSO indices (Dp) are constructed for lead times up to 48 months to incorporate the combined influence of all potential tropical oceanic teleconnections associated with ISMR. First, we obtain the correlation maps between global tropical D20 and ISMR anomalies over a long time period. The D20-ISMR correlation maps constructed at different lead times represent the monthly evolution of ENSO in the Pacific Ocean and the associated teleconnection patterns in the Atlantic and Indian Oceans. Next, for each lead, the index is obtained by projecting the D20 anomalies onto the statistically significant regions of the long-term correlation map between D20 and ISMR anomalies. For example, the 18-month lead G-ENSO index (Dp(-18)) is obtained by projecting the December D20 anomaly field over the global tropics for the period 1873-2008 onto regions that exhibit statistically significant (≥95%) correlation between December D20 anomalies (1873-2008) and ISMR anomalies (1875-2010). The projection is computed through a Hadamard (element-wise) multiplication between the D20 anomaly field and the correlation map, followed by spatial summation over all grid points as shown in Figure 1 of (Sharma et al., 2022). A schematic representation and detailed description of the methodology are provided in Supplementary Figure S1 and Text S1, respectively.

### 2.3 Modelling Two Interacting Climate Systems

Descriptors of the ISMR and the G-ENSO can be quantified by averaged order parameters arising from the interaction of two spatially extended oscillatory media. Let $A(x,y,t)$ and $B(x,y,t)$ denote complex-valued order parameters whose spatiotemporal evolution is governed by two coupled CGLEs

$$\partial_t A = A + (1 + ia_1)\nabla^2 A - (1 + ia_3)|A|^2 A - \gamma B \quad (1)$$

$$\kappa\partial_t B = B + (1 + ib_1)\nabla^2 B - (1 + ib_3)|B|^2 B + \gamma A \quad (2)$$

We define the *theoretical* ISMR (*I*) and G-ENSO (*E*) descriptors as

$$I(t) = \frac{I_0}{N^2} \iint \mathrm{d}x\mathrm{d}y \ \mathrm{Re}A \quad (3)$$

$$E(t) = \frac{E_0}{N^2} \iint \mathrm{d}x\mathrm{d}y \ \mathrm{Re}B \quad (4)$$

Here $N$ is the spatial resolution (we have taken $N = 256$ in all simulations) and Re denotes the real part. The dimensionless constants $E_0 = I_0 = 25$ are fixed to match the window of the experimentally observed amplitudes. $A$ and $B$ depend on time $t$ and on two spatial coordinates $x$, $y$. The Laplacian $\nabla^2$ operator introduces a diffusive (local) coupling within each oscillatory media ('weather' interactions). The real parameters $a_1$ and $b_1$ characterize linear dispersion while parameters $a_3$ and $b_3$ describe nonlinear saturation. Parameter $k$ sets the time scale of $B$ compared to the one of $A$. For $k > 1$ ($k < 1$), $B$ evolves slower (faster) than $A$. Since the "oceanic" G-ENSO is slower than the "atmospheric" ISMR, we shall take $k = 2$ in the simulations.

The coupling constant $\gamma$ describes the in-phase amplification of $B$ by $A$ and the weakening of $A$ by $B$: Regions with high $A$ tend to produce regions with high in-phase $B$ which, in turn, weaken high-$A$ regions. The resulting lower-$A$ regions receive then an increased $A$ through diffusion from neighboring high-$A$ regions, creating a time delay between the in-phase synchrony of $A$ and $B$.

These *locally linear* dynamical balances are at the very heart of our *nonlinear* model. The resulting cycles exhibit complex behavior because the spatial regions migrate fueled by the nonlinear instability present in the system. The latter creates incoherent (i.e. turbulent) patterns: The parameters $a_1$, $a_3$, $b_1$, $b_3$ are selected so that the individual media are both in the Benjamin-Feir unstable regime $1+a_1a_3 < 0$ and $1+b_1b_3 < 0$.

## 3 Results

### 3.1 Estimate of LDP from observations

The ISMR is the summer phase of annual cycle of rainfall over the Indian land mass (Figure S3). The ISMR oscillator time series is represented by the 5-month moving average of the monthly mean rainfall time series spatially averaged over the Indian land mass (Figure 1c) while the G-ENSO is represented by the first Principal Component of monthly mean D20 over the tropics between $30^{O}$S and $30^{O}$N (Figure 1d) with its spatial structure represented by the EOF1 (Figure 1a). The high correlation between ISMR and G-ENSO happens at a lag of 18-month and is shown in Figure 1b. There, the ISMR curve is the June to September mean rainfall over India and the Dp curve is a G-ENSO based predictor of ISMR, derived from projection of D20 at 18-month lag (Sharma et al., 2022).

To get an estimate of the LDP of ISMR, we calculate the growth of errors from the observed ISMR time series (Figure1c) following a method described in (Goswami & Xavier, 2003) (Supplementary text and Figure S2). At the peak of the monsoon season, the initial error is very high due to large event-to-event variability but the error decreases with lead time and has a minimum in the winter and increases again peaking in the next summer (Figure 2a). A strong locking of ISMR with the annual cycle makes growth of errors also follow an annual cycle. Thus, the growth of errors is a function of initial conditions coming from different calendar months giving an LDP of ISMR of about 6 months with best forecasts coming from the previous winter.

The growth rate of errors in ISMR with lead time (Figure 2a) shows that there are minima of error growth at 19-months and 31-months with potential predictability at those leads. A similar error growth estimate from the peaks of the G-ENSO oscillator (Figure 1d) indicates that initial conditions starting from the peak (El-Niño) also has large initial error (Figure 2b) and is limited by the 'fast' growth of errors (Goswami & Shukla, 1991) and saturate quickly limiting predictability up to only about 10 months, consistent with the skill of most dynamical prediction systems (Ham et al., 2019). However, for initial conditions starting from the trough of the oscillation (La-Niña), the initial error is relatively small, and the error grows following the 'slower' of the two modes of growth, limiting potential ENSO predictability up to approximately 30-months (Goswami & Shukla, 1991) (Figure 2b). However, dynamical models may not be able to harness this predictability as the initial error is also a function of the annual cycle with large errors (event to event variability) during certain seasons (e.g., the spring predictability barrier) (Figure 2c-e). In contrast, data driven deep learning models use advanced 'signal processing' and have demonstrated that they may be able to overcome the error growth issue and harness such predictability up to 18-month lead (Ham et al., 2019). Due to paucity of oceanic sub-surface observations, the 'initial error' in the coupled seasonal prediction systems is in the nonlinear regime. It is notable that due to the locking of both ISMR and G-ENSO with the annual cycle, there is a minimum in the forecast errors starting from the peak at approximately 19-month lead (Figure 2b). The strong correlation between the two systems at 18-month lead is consistent with minimum in growth of errors in both oscillators at 19-month lead.

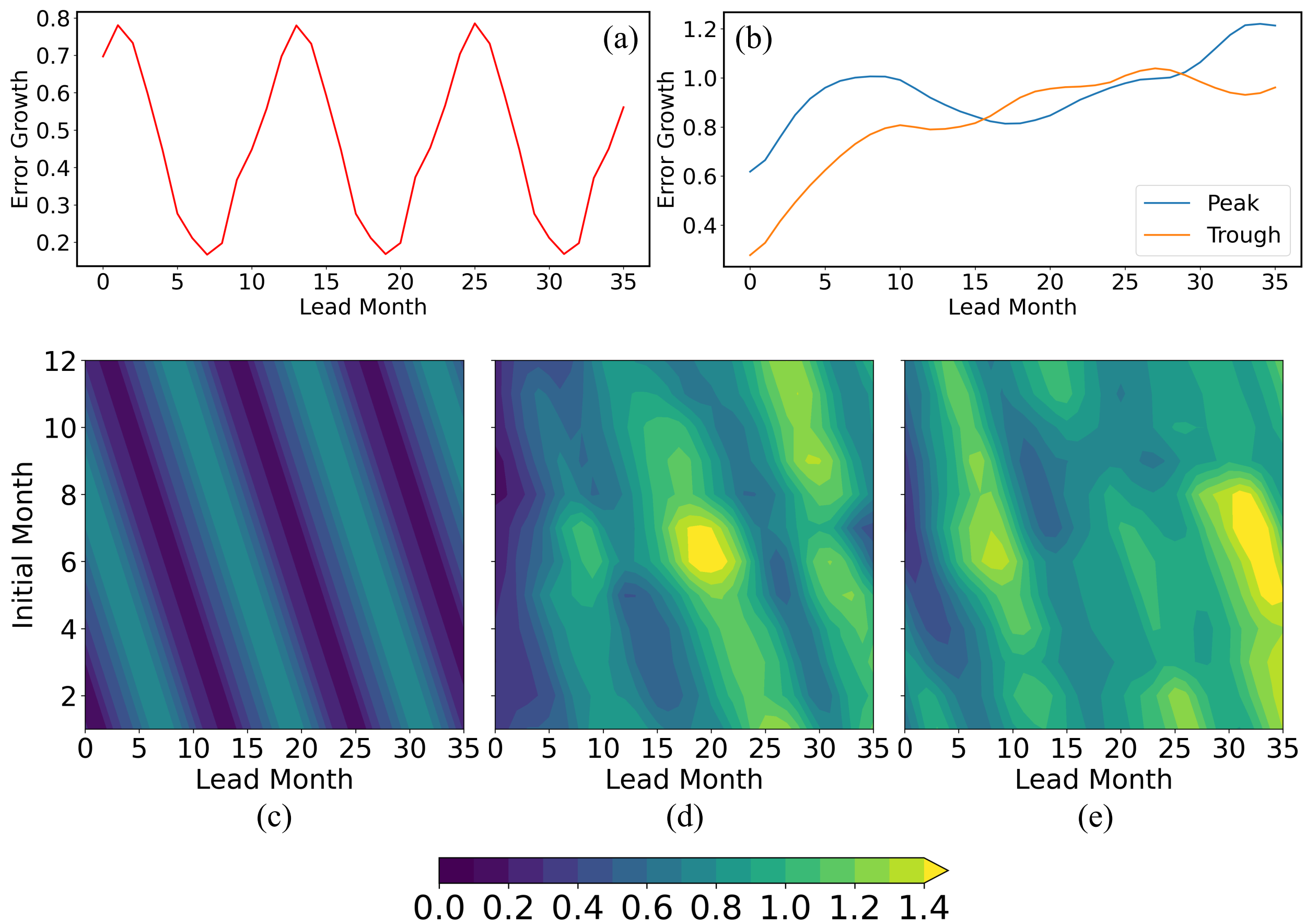

**Figure 2: Growth of errors in ISMR and G-ENSO. (a)** Growth rate of errors in ISMR as a function of lead forecast month estimated from a 5-month window time series of ISMR following the method described in the Supplementary Text S2. Since the error growth here is dominated by an annual cycle, we present the growth of errors starting from the peak monsoon season only. **(b)** Same as (a) but for the G-ENSO time series (Figure 1d) and starting from the peak and trough of the oscillation. **(c)** Growth of errors of ISMR as a function of lead month and as a function of initial conditions from all calendar months. **(d)** Same as (c) but for G-ENSO time series starting from trough (<-1, Figure 1d). The dominant modulation by the annual cycle is evident. **(e)** Same as (c) but for G-ENSO time series starting from peak (>1, Figure 1d). The dominant modulation by the annual cycle is evident.

### 3.2 Lag synchronization between ISMR and G-ENSO

Order parameters $A$ and $B$ are initialized with different random uniform noises and integration of Eqs. (1) and (2) is carried over 4000 time steps, each of size $\Delta T = 0.05$ by using a pseudoespectral method implementing an exponential time-stepping algorithm (Cox & Matthews, 2002). We set $t = 0$ for the generic situation reached after eliminating the transient behavior contained in the first 1600 time steps. Snapshots of the spatial arrangement at $t = 0$ of $A$ and $B$ are shown in Figure 3a (left) and 3b respectively. There is no obvious apparent relationship between these patterns. However, the spatial arrangement of $A$ after 18 months shown in Figure 3a (right) shows strong similarities with the snapshot of $B$ in Figure 3b, which can act as a predicted pattern for $A$. These similarities arise because of the impact of the coupling between $A$ and $B$ in their self-organization. This coupling, caused by terms $\gamma A$ and $-\gamma B$, is stronger in regions where the amplitudes are large (light regions). Because of it, $A$ and $B$ are strongly correlated with 18-month lag, keeping this good correlation for any arbitrary long time and making it predictable.

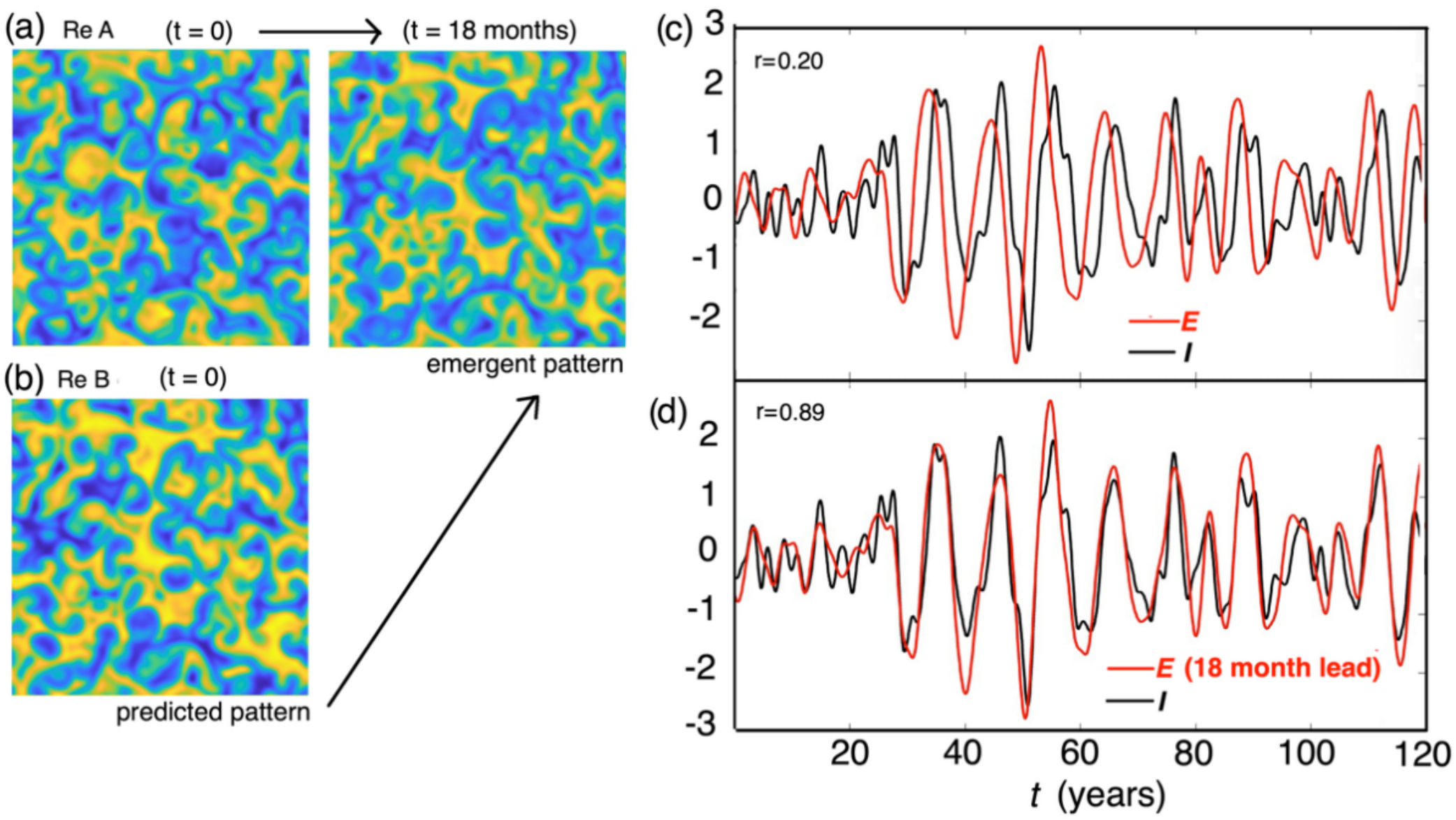

**Figure 3: Lag synchronization between non-linear aperiodic oscillators. (a)** Spatial snapshots of the real part of order parameter $A$ at $t = 0$ (left) and $t = 18$ months (right). **(b)** Spatial snapshot of $B$ at $t = 0$; in both (a) and (b) high amplitude values are yellow (light) and low amplitude values are blue (dark). The transient behavior before $t = 0$ (starting from uniformly random initial conditions) has been eliminated. The pattern of $B$ in (b) is strongly correlated to the pattern of $A$ that emerges from Eqs. (1) and (2) after 18 months and can be used as a predictor. **(c)** Time series for $I$ (black curve) and $E$ (red curve) obtained from Eqs. (3) and (4) respectively. **(d)** The same as in (d) but with $E$ shifted 18 months to the future. Parameter values: $a_1 = b_1 = -0.4$; $a_3 = 3.0$, $b_3 = 2.5$, $k = 2$, $\gamma = 0.8$.

The order parameters $A$ and $B$ are internal degrees of freedom of the averaged variables $I$ and $E$, respectively. Incoherent patterns in $A$ or $B$, lead to aperiodic, chaotic-like oscillations in $I$ and $E$, respectively. In Figure 3c the time series of $I$ and $E$ are shown. The time lag with which $E$ anticipates $I$ is apparent. Although the 0-lag correlation in Figure 3c is poor ($r = 0.2$), by shifting $E$ 18 months to the future, excellent 18 month-lag correlation ($r = 0.89$) is found. This implies that *E can be used to predict I with 18 months in advance*. Figure 3d is in excellent qualitative agreement with the observation from meteorological data in Figure 1b.

A sufficiently large coupling constant $\gamma$ is necessary to achieve lag synchronization. By keeping all other parameters as in Figure 3, we can study the impact of changing $\gamma$ in the lag to which maximum correlation happens. We show in Figure 4 the correlation as a function of all leads from 0 to 30 months, as obtained from Eqs. (1) to (4). When $\gamma = 0$, $A$ and $B$ are independent and, therefore, $I$ and $E$ are uncorrelated. When $\gamma$ is increased to a significant value above zero, a lead for which the correlation is maximum appears. The maximum correlation increases with increasing $\gamma$. When $\gamma = 0.8$, the maximum correlation 0.89 is found at 18 months lead. This red curve in Figure 4 compares qualitatively well with the red curve found for Dp in Figure 2c in (Sharma et al., 2022).

A lead-lag correlation study between ISMR (assuming July as center) and monthly Niño 3.4 SST (Figure S4c) shows that maximum negative correlation between the two occurs about three months after the peak monsoon. This is likely to be a result of positive feedback between the two. An evolving El Niño during June-September results in a weakening of the monsoon rainfall (in our model, $B$ weakens $A$) and an anomalous negative monsoon heat source and anomalous westerly wind anomaly over equatorial western Pacific through the Gill response (Gill, 1980). Enhancement of the westerly wind bursts in the western Pacific further amplifies the El Niño (i.e. $A$ enhances $B$). During La Niña, the situation simply reverses and a stronger monsoon rainfall further strengthens La Niña. These observations form the basis for the assumption of strong antisymmetric bi-directional coupling in our model.

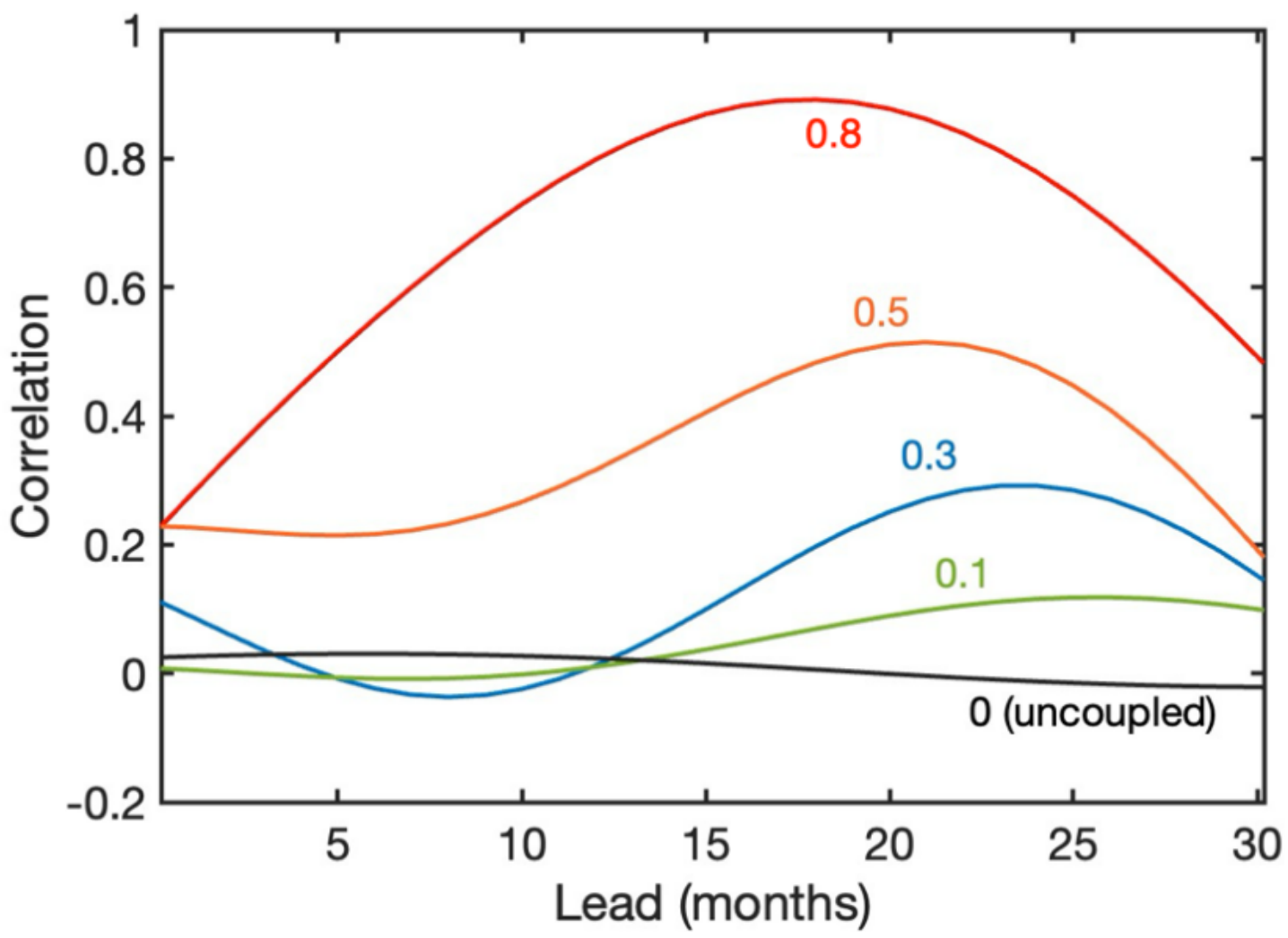

**Figure 4:** Correlation coefficient $r$ between $I$ and $E$ as a function of lead and coupling strength. Each curve is obtained from Eqs. (1) to (4) by fixing the value of $\gamma$ indicated close to the curve and shifting the time series of $E$ to the future according to the lead values. All other parameter values are as in Figure 3.

## 4 Conclusions

In this letter we have shown that lag synchronization between interacting climate systems (viewed as nonlinear spatially extended oscillatory media) is the dynamical basis behind the experimentally observed high potential predictability of the ISMR from the G-ENSO at 18-month lead (Sharma et al., 2022). Appropriate complex-valued order parameters characterize the internal degrees of freedom of different climate systems. The dynamical impact of the antisymmetric bidirectional coupling of the order parameters can be summarized as follows: *The coupling leads to the internal self-organization of the degrees of freedom within each climate system so that lag synchronization arises between different climate systems as an emergent phenomenon.* If the coupling constants were zero, predictability beyond the LDP would not be possible.

We have shown the relevance of lag synchronization for weather and climate science and its potential for long-lead climate predictability. Our findings have far-reaching implications in long-lead seasonal prediction to other regional climates driven by the G-ENSO as well (not shown here). Our preliminary investigation indicates that extra-tropical climates such as the NAO could be more predictable than previously thought. So far, seasonal predictability is still being looked within the LDP of the climate systems (Charney & Shukla, 1981; Shukla, 1981, 1998). We have unraveled here a novel source of long-lead predictability beyond the LDP with help of a model that is able to capture the high correlation observed in the observation 'experiments'. Our findings pave the way for data driven models to make useful forecasts of seasonal climate at long-leads, thought impossible previously.

**Acknowledgements:** VGM gratefully acknowledges a Visiting Faculty Fellowship in the Department of Aerospace Engineering at IIT Madras under the Institute of Eminence (IoE) program. BNG thanks Anusandhan National Research Foundation, Government of India for the ANRF Prime Minister Professorship and Gauhati University for support. DS and RIS acknowledge the funding from IndusInd Bank through the CSR project (CR23242519AEINIB002696) and the IoE initiative (Grant No. SP22231222CPETWOCTSHOC).

**Funding:** This project was supported by IndusInd Bank through a CSR project (CR23242519AEINIB002696) and the Institute of Eminence (IoE) initiative of the Ministry of Human Resources and Development (MHRD) at IIT Madras; project number: SP22231222CPETWOCTSHOC.

**Availability Statement:** Data related to this paper can be downloaded from:

SODA version 2.2.4 (for D20) http://apdrc.soest.hawaii.edu/datadoc/soda_2.2.4.php
IITM ISMR https://tropmet.res.in/static_pages.php?page_id=53
COBE SST2 https://psl.noaa.gov/data/gridded/data.cobe2.html
ERA5 ISMR https://cds.climate.copernicus.eu/datasets/reanalysis-era5-single-levels-monthly-means?tab=download

**Competing interests:** Authors declare that they have no competing interests.

**Author contributions:**

Conceptualization: VGM, DS, ST, BNG, RIS
Methodology: VGM, DS, BNG
Investigation: VGM, DS, ST
Visualization: VGM, DS, ST
Supervision: BNG, RIS
Writing – original draft: VGM, BNG, RIS
Writing – review & editing: All authors

**References:**

Andrew Moore, B. M., & Kleeman, R. (1996). The dynamics of error growth and predictability in a coupled model of ENSO. *Quarterly Journal of the Royal Meteorological Society*, *122*, 1405–1446.

Aranson, I. S., & Kramer, L. (2002). The world of the complex Ginzburg-Landau equation. *Reviews of Modern Physics*, *74*(1), 99.

Battisti, D. S., Hirst, A. C., & Sarachik, E. S. (1989). Instability and predictability in coupled atmosphere-ocean models. *Phil. Trans. R. Soc. Lond. A*, *329*, 237–247. Retrieved from http://rsta.royalsocietypublishing.org/Downloadedfrom

Bjerknes, J. (1969). Atmospheric Teleconnections From the Equatorial Pacific. *Monthly Weather Review*, *97*(3), 163–172. Retrieved from http://journals.ametsoc.org/doi/abs/10.1175/1520-0493(1969)097%3C0163:ATFTEP%3E2.3.CO;2

Borah, P. J., Venugopal, V., Sukhatme, J., Muddebihal, P., & Goswami, B. N. (2020). Indian monsoon derailed by a North Atlantic wavetrain. *Science*, *370*(6522), 1335–1338. https://doi.org/10.1126/science.aay6043

Carton, J. A., & Giese, B. S. (2008). A Reanalysis of Ocean Climate using Simple Ocean Data Assimilation (SODA). *Monthly Weather Review*, *136*(8), 2999–3017. https://doi.org/10.1175/2007MWR1978.1

Carton, J. A., Chepurin, G. A., & Chen, L. (2018). SODA3: A new ocean climate reanalysis. *Journal of Climate*, *31*(17), 6967–6983. https://doi.org/10.1175/jcli-d-18-0149.1

Charney, J. G., & Shukla, J. (1981). Predictability of monsoons. In J. Lighthill & R. Pearce (Eds.), *Monsoon dynamics* (pp. 99–109). Cambridge University Press. https://doi.org/10.1017/cbo9780511897580.009

Ciszak, M., Mayol, C., Mirasso, C. R., & Toral, R. (2015). Anticipated synchronization in coupled complex Ginzburg-Landau systems. *Physical Review E - Statistical, Nonlinear, and Soft Matter Physics*, *92*(3). https://doi.org/10.1103/PhysRevE.92.032911

Cox, S. M., & Matthews, P. C. (2002). Exponential time differencing for stiff systems. *Journal of Computational Physics*, *176*(2), 430–455. https://doi.org/10.1006/jcph.2002.6995

Cross, M. C., & Hohenberg, P. C. (1993). Pattern formation outside of equilibrium. *Reviews of Modern Physics*, *65*, 851.

Duane, G S, Tribbia, J. J., & Weiss, J. B. (2006). *Nonlinear Processes in Geophysics Synchronicity in predictive modelling: a new view of data assimilation*. *Nonlin. Processes Geophys* (Vol. 13). Retrieved from www.nonlin-processes-geophys.net/13/601/2006/

Duane, Gregory S. (1997). Synchronized chaos in extended systems and meteorological teleconnections. *Physical Review E*, *56*(6), 6475–6493.

Duane, Gregory S. (2015). Synchronicity from synchronized chaos. *Entropy*, *17*(4), 1701–1733. https://doi.org/10.3390/e17041701

Duane, Gregory S., & Tribbia, J. J. (2001). Synchronized chaos in geophysical fluid dynamics. *Physical Review Letters*, *86*(19), 4298–4301. https://doi.org/10.1103/PhysRevLett.86.4298

Duane, Gregory S, & Tribbia, J. J. (2004). Weak Atlantic-Pacific Teleconnections as Synchronized Chaos. *Journal of the Atmospheric Sciences*, *61*, 2149–2797.

Duane, Gregory S, Webster, P. J., & Weiss, J. B. (1999). Co-occurrence of Northern and Southern Hemisphere Blocks as Partially Synchronized Chaos. *Journal of the Atmospheric Sciences*, *56*, 4183–4205.

García-Morales, V., & Krischer, K. (2012). The complex Ginzburg-Landau equation: An introduction. *Contemporary Physics*, *53*(2), 79–95. https://doi.org/10.1080/00107514.2011.642554

García-Morales, V., Tandon, S., Kurths, J., & Sujith, R. I. (2024). Universality of oscillatory instabilities in fluid mechanical systems. *New Journal of Physics*, *26*(3). https://doi.org/10.1088/1367-2630/ad2bb1

Giese, B. S., & Ray, S. (2011). El Niño variability in simple ocean data assimilation (SODA), 1871-2008. *Journal of Geophysical Research: Oceans*, *116*(2), 1–17. https://doi.org/10.1029/2010JC006695

Gill, A. E. (1980). Some simple solutions for heat-induced tropical circulation. *Quarterly Journal of the Royal Meteorological Society*, *106*, 447–462. https://doi.org/https://doi.org/10.1002/qj.49710644905

Goswami, B. N., & J.Shukla. (1991). Predictability of a Coupled Ocean-Atmosphere Model. *Journal of Climate*, *4*, 3–22. Retrieved from https://doi.org/10.1175/1520-390 0442(1991)004%3C0003:POACOA%3E2.0.CO;2

Goswami, B. N., & Shukla, J. (1991). Predictability of a Coupled Ocean-Atmospheric Model Goswami Shukla. *Journal of Climate*, *4*, 3–22.

Goswami, B. N., & Xavier, P. K. (2003). Potential predictability and extended range prediction of Indian summer monsoon breaks. *Geophysical Research Letters*, *30*(18), 1–4. https://doi.org/10.1029/2003GL017810

Goswami, B. N., Madhusoodanan, M. S., Neema, C. P., & Sengupta, D. (2006). A physical mechanism for North Atlantic SST influence on the Indian summer monsoon. *Geophysical Research Letters*, *33*, L02706. https://doi.org/10.1029/2005GL024803

Griffies, S. M., & Bryan, K. (1997). A predictability study of simulated North Atlantic multidecadal variability. *Climate Dynamics*, *13*, 459–487.

Ham, Y. G., Kim, J. H., & Luo, J. J. (2019). Deep learning for multi-year ENSO forecasts. *Nature*, *573*(7775), 568–572. https://doi.org/10.1038/s41586-019-1559-7

Hirahara, S., Ishii, M., & Fukuda, Y. (2014). Centennial-Scale Sea Surface Temperature Analysis and Its Uncertainty. *Journal of Climate*, *27*(1), 57–75. https://doi.org/10.1175/JCLI-D-12-00837.1

Jiang, J., Huang, Z. G., Grebogi, C., & Lai, Y. C. (2022). Predicting extreme events from data using deep machine learning: When and where. *Physical Review Research*, *4*(2), 023028.

Keane, R. J., Parker, D. J., Dunn-Sigouin, E., Kolstad, E. W., & Marsham, J. H. (2025). Mid-Latitude Versus Tropical Scales of Predictability and Their Implications for Forecasting. *Meteorological Applications*, *32*(4), 1–14. https://doi.org/10.1002/met.70055

Kirtman', B. P., & Shukla, J. (2000). Influence of the Indian summer monsoon on ENSO. *Quarterly Journal of the Royal Meteorological Society*, *126*, 213–239.

Kocarev, L., & Parlitz, U. (1996). Generalized Synchronization, Predictability, and Equivalence of Unidirectionally Coupled Dynamical Systems. *Physical Review Letters*, *76*(11), 1816–1819.

Latif, M. (1998). Dynamics of Interdecadal Variability in Coupled Ocean-Atmosphere Models. *Journal of Climate*, *11*, 602–624.

Lorenz, E. N. (1963). Deterministic Nonperiodic Flow. *Journal of the Atmospheric Sciences*, *20*, 130–141.

Lorenz, E. N. (1969). The predictability of a flow which possesses many scales of motion. *Tellus*, *21*(3), 289–307. https://doi.org/10.1111/j.2153-3490.1969.tb00444.x

M. Benno Blumenthal. (1991). Predictibility of a Coupled Ocean-Atmosphere Model. *Journal of Climate*, *4*(8), 766–784. Retrieved from https://doi.org/10.1175/1520-0442(1991)004%3C0766:POACOM%3E2.0.CO;2

Monroy, D. L., & Naumis, G. G. (2021). Description of mesoscale pattern formation in shallow convective cloud fields by using time-dependent Ginzburg-Landau and Swift- Hohenberg stochastic equations. *Physical Review E*, *103*(3), 032312.

Parthasarathy, B., Munot, A. A., & Kothawale, D. R. (1994). All-India monthly and seasonal rainfall series: 1871-1993. *Theoretical and Applied Climatology*, *49*(4), 217–224. https://doi.org/10.1007/BF00867461

Pecora, L. M., & Carroll, T. L. (1990). Synchronization in Chaotic Systems. *Physical Review Letters*, *64*(8), 821–824.

Peixoto, J. P., & Oort, A. H. (1984). Physics of climate. *Reviews of Modern Physics*, *56*, 365–426.

Philander, S. G. H. (1983). El Nino Southern Oscillation Phenomena. *Nature*, *302*(5906), 295–301.

Pikovsky, A., Rosenblum, M., & Kurths, J. (2001). *Synchronization: A universal concept in nonlinear sciences*. (B. Chirikov, F. Moss, P. Cvitanovi´, & H. Swinney, Eds.). Cambridge University Press.

Rajesh, P. V., & Goswami, B. N. (2020). Four-dimensional structure and sub-seasonal regulation of the Indian summer monsoon multi-decadal mode. *Climate Dynamics*, *55*(9–10), 2645–2666. https://doi.org/10.1007/s00382-020-05407-y

Rasmusson, E. M., & Carpenter, T. H. (1983). The Relationship Between Eastern Equatorial Pacific Sea Surface Temperatures and Rainfall over India and Sri Lanka. *Monthly Weather Review*, *111*, 517–528. https://doi.org/https://doi.org/10.1175/1520-0493(1983)111<0517:TRBEEP>2.0.CO;2

Rosenblum, M. G., Pikovsky, A. S., & Kurths, J. (1996). Phase Synchronization of Chaotic Oscillators. *Physical Review Letters*, *76*(11), 1804–1807.

Rosenblum, M. G., Pikovsky, A. S., & Kurths, J. (1997). From Phase to Lag Synchronization in Coupled Chaotic Oscillators. *Physical Review Letters*, *78*(22), 4193–4196.

Sharma, D., Das, S., Saha, S. K., & Goswami, B. N. (2022). Mechanism for high "potential skill" of Indian summer monsoon rainfall prediction up to two years in advance. *Quarterly Journal of the Royal Meteorological Society*, *148*(749), 3591–3603. https://doi.org/10.1002/qj.4375

Sharma, D., Das, S., Chakraborty, D., Mitra, A., & Goswami, B. N. (2025). Improving Indian summer monsoon rainfall prediction using deep learning up to two years in advance. *Quarterly Journal of the Royal Meteorological Society*, *152*(774), e70023. https://doi.org/10.1002/qj.70023

Shukla, J. (1981). Dynamical predictability of monthly means. *Journal of the Atmospheric Sciences*, *38*(12), 2547–2572. https://doi.org/10.1175/1520-0469(1981)038<2547:DPOMM>2.0.CO;2

Shukla, J. (1998). Predictability in the Midst of Chaos: A Scientific Basis for Climate Forecasting. *Science*, *282*(5389), 728–731.

Shukla, J., & Paolino, D. A. (1983). The Southern Oscillation and long-range forecasting of the summer monsoon rainfall over India. *Monthly Weather Review*, *111*(9), 1830–1837. https://doi.org/10.1175/1520-0493(1983)111<1830:TSOALR>2.0.CO;2

Timmermann, A., An, S. Il, Kug, J. S., Jin, F. F., Cai, W., Capotondi, A., et al. (2018). El Niño–Southern Oscillation complexity. *Nature*, *559*(7715), 535–545. https://doi.org/10.1038/s41586-018-0252-6

Walker, G. T. (1924). Correlation in Seasonal Variations of Weather, IX. A Further Study of World Weather. *Memoirs of the India Meteorological Department*, *24*(9), 275–333.

Wallace, J. M. ., Battisti, D. S. ., Thompson, D. W. J. ., & Hartmann, D. L. . (2023). *The atmospheric general circulation*. Cambridge University Press.

Wang, C., Deser, C., Yu, J. Y., DiNezio, P., & Clement, A. (2017). El Niño and Southern Oscillation (ENSO): A Review. In *Coral Reefs of the World* (Vol. 8, pp. 85–106). Springer Nature. https://doi.org/10.1007/978-94-017-7499-4_4

Webster, P. J. (1995). The Annual Cycle and the Predictability of the Tropical Coupled Ocean-Atmosphere System. *Meteorol. Atmos. Phys*, *56*, 33–55.

Webster, P. J., Magaña, V. O., Palmer, T. N., Shukla, J., Tomas, R. A., Yanai, M., & Yasunari, T. (1998). Monsoons: processes, predictability, and the prospects for prediction. *Journal of Geophysical Research: Oceans*, *103*(C7), 14451–14510. https://doi.org/10.1029/97jc02719

Yang, S.-C., Baker, D., Li, H., Cordes, K., Huff, M., Nagpal, G., et al. (2006). Data Assimilation as Synchronization of Truth and Model: Experiments with the Three-Variable Lorenz System*. *Journal of the Atmospheric Sciences*, *63*, 2340–2354.

Zhou, C. T. (2006). Synchronization in nonidentical complex Ginzburg-Landau equations. *Chaos*, *16*(1), 013124. https://doi.org/10.1063/1.2170459

# Supplementary Materials for

## Why Seasonal Climate is Predictable Beyond the limit of Deterministic Climate Chaos?

Vladimir García-Morales[1,2], Devabrat Sharma[2,3], Shruti Tandon[2,3] , B. N. Goswami[4], and R. I. Sujith[2,3]

[1]Departament de Física de la Terra i Termodin`amica, Universitat de València, E-46100 Burjassot, Spain
[2]Department of Aerospace Engineering, Indian Institute of Technology Madras, Chennai- 600036, India
[3]Centre of Excellence for Studying Critical Transitions in Complex Systems, Indian Institute of Technology Madras, Chennai-600036, India
[4]ST Radar Centre, Gauhati University, Guwahati-781014, India

*Corresponding author. Email:

**Content of this file:**

Text S1-S2
Figure S1-S4

## Text S1: Global-ENSO indices

The Global-ENSO index Dp at a given lead month is constructed to incorporate the combined influence of all potential tropical oceanic teleconnections associated with Indian Summer Monsoon Rainfall (ISMR) at that lead (Sharma et al., 2022).

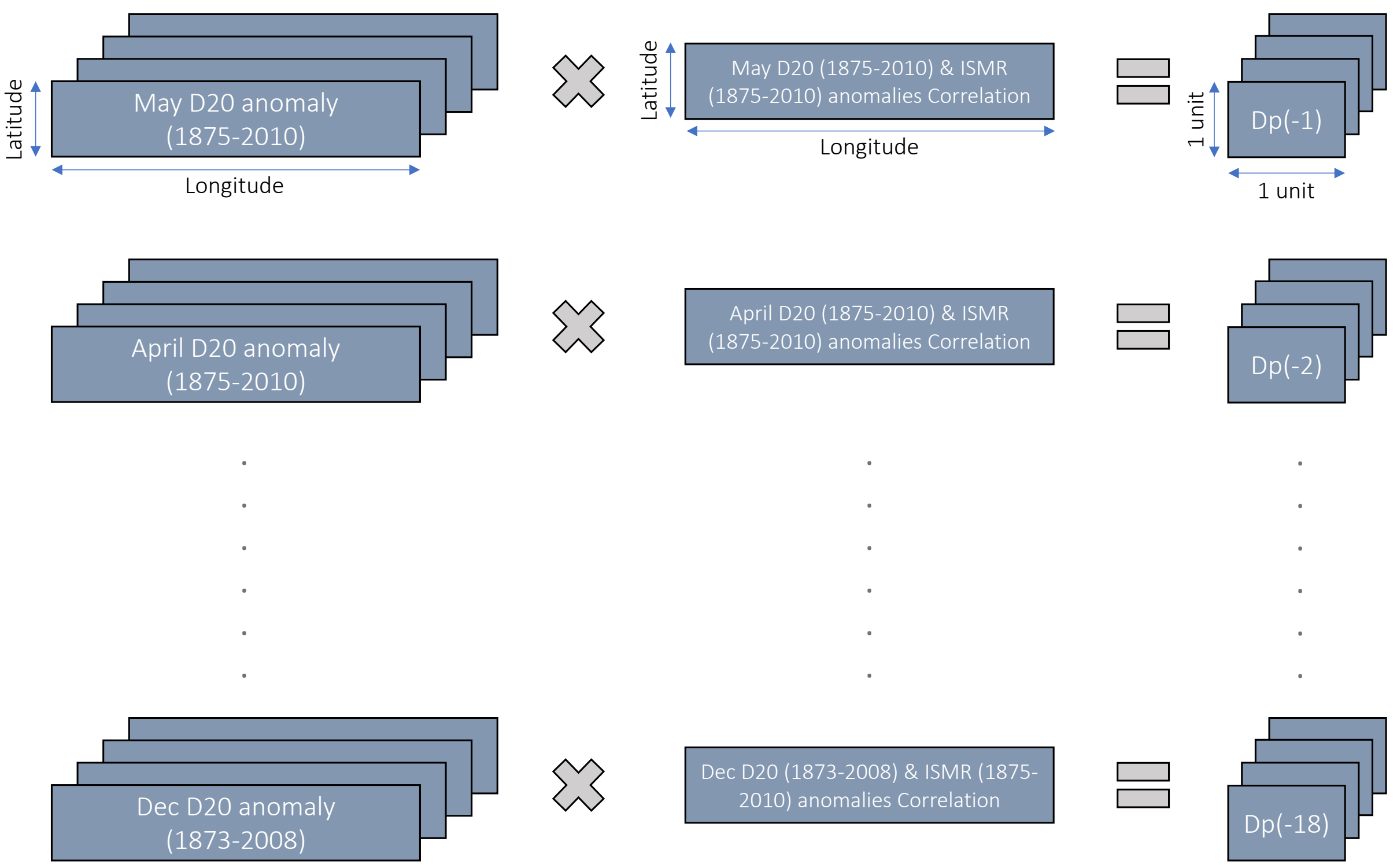

**Figure S1:** Schematic representation of the algorithm developed by (Sharma et al., 2022) for generating D20-based Global-ENSO indices.

For example, Dp at a one-month lead is constructed using global tropical May D20 anomalies over 0°–360°E and 30°S–30°N from the Simple Ocean Data Assimilation (SODA) version 2.2.4 dataset for the period 1875–2010. These anomalies are multiplied element-wise by the statistically significant regions (95% confidence level) of the correlation map between global tropical D20 anomalies and ISMR anomalies from (Parthasarathy et al., 1994) for the same period (1875–2010). The resulting values are then summed over all grid boxes to obtain a single time series, referred to as Dp at one-month lead or Dp(-1).

Similarly, Dp at an 18-month lead is constructed using global tropical December D20 anomalies over 0°–360°E and 30°S–30°N from SODA version 2.2.4 for the period 1873–2008. These anomalies are multiplied element-wise by the correlation map between global tropical D20 anomalies (1873–2008) and ISMR anomalies from (Parthasarathy et al., 1994) for the period 1875–2010. The resulting grid-box values are then summed to produce the Dp(-18) time series.

**Text S2: Growth of Errors**

The growth of error from different initial conditions (January–December) was calculated for both rainfall and the first principal component (PC1) of the dominant empirical orthogonal function (EOF) of global tropical monthly D20 anomalies, following the method of (Goswami & Xavier, 2003). In this approach, the standard deviation in the evolution of the system from a given initial condition is used as a proxy for error growth.

For example, in the case of 5-month running mean rainfall anomalies (Figure S2a and 1c), the time step at which July lies at the center of the running mean is taken as the initial condition (lead 0) (Figure S2c). The standard deviation of all rainfall anomaly values corresponding to this July-centered time steps is defined as the error at lead 0 (Figure S2g). The same procedure is then repeated for subsequent time steps: the standard deviation corresponding to the August-centered running mean gives the error at lead 1, the September-centered running mean gives the error at lead 2, and so on (Figure S2g and 2a). This procedure provides the evolution of error growth in the 5-month running mean rainfall anomalies for the June initial condition.

The same analysis was repeated by individually considering January- to December-centered time steps as the initial condition (lead 0). This resulted in a matrix of error growth (Figure 2c), which represents the evolution of error in the 5-month running mean rainfall anomaly for different initial conditions throughout the annual cycle.

For the PC1 of the dominant EOF of global tropical monthly D20 anomalies (Figure S2b and 1d), time steps with PC1 values greater than 1 (less than −1) were considered as the initial conditions (lead 0) for peak (trough) events (Figure S2e-f). The standard deviation of these selected time steps was defined as the error at lead 0. Subsequently, the standard deviation of all time steps one month ahead of the selected events was calculated as the error at lead 1, and the procedure was repeated for successive months. Using this approach, the error growth associated with strong positive (negative) D20 events was estimated up to 36-month leads (Figure S2h and 2b).

Further, the analysis was repeated separately for events occurring during January to December by individually considering each calendar month as the initial condition. This resulted in a matrix of error growth (Figure 2d-e), which represents the evolution of error growth up to 36-month leads in global tropical monthly D20 anomalies for strong positive (negative) events initiated in different months of the year.

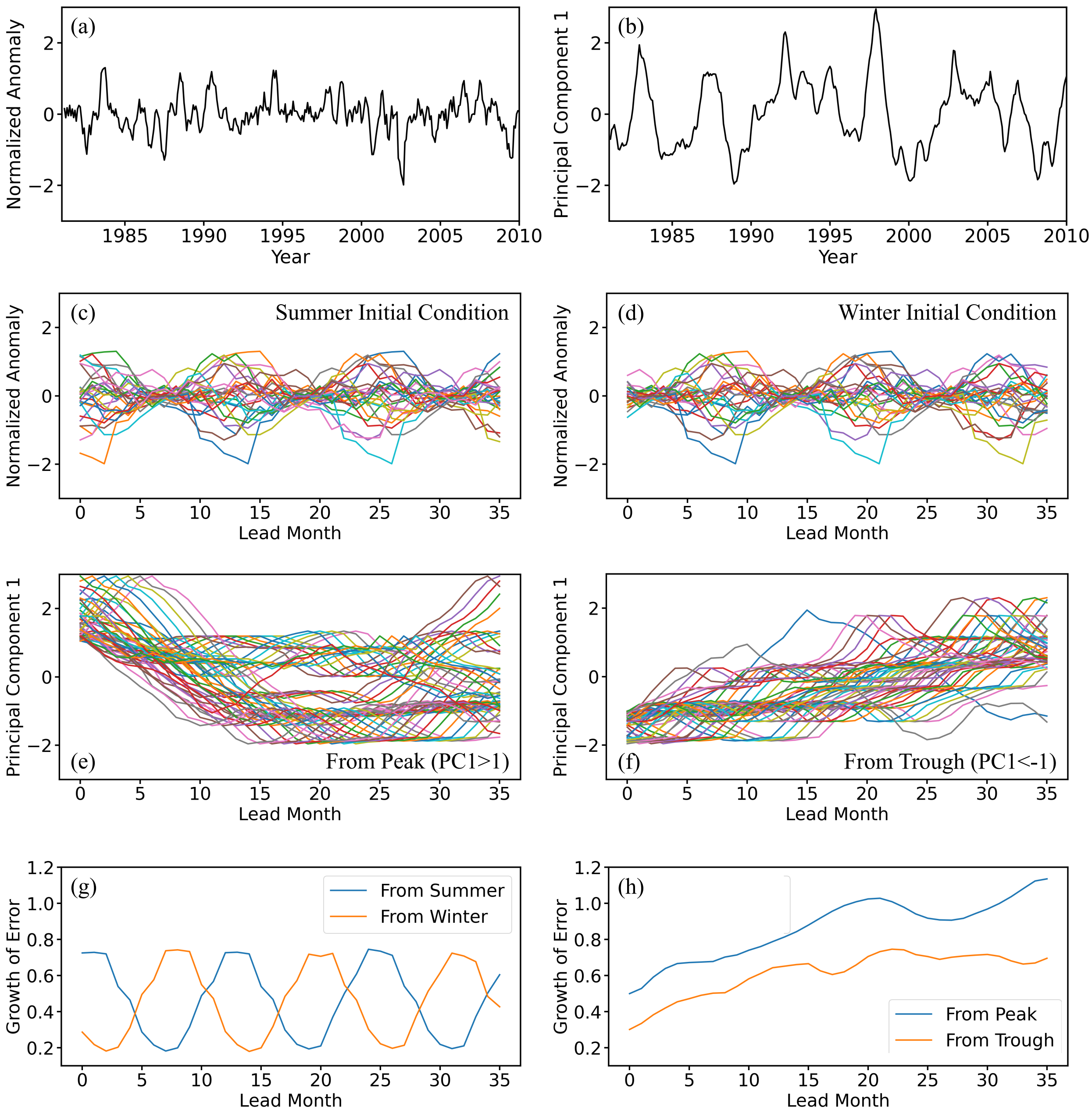

**Figure S2: (a)** Time series of 5-month running mean of ISMR annual cycle for 30-years, **(b)** Same as (a) but for G-ENSO (PC1) for 30-years. **(c)** Spaghetti diagram of evolution of evolution of ISMR from the all peaks (summer), **(d)** same as (c) but for all troughs (winter), **(e)** Spaghetti diagram of evolution of evolution of G-ENSO from the all peaks, **(f)** same as (e) but for all troughs. **(g)** Growth of errors as a function of lead time for ISMR starting from peaks and troughs based on the 30-year data shown in (a), **(h)** Same as (f) but for G-ENSO based on 30-year data shown in (b).

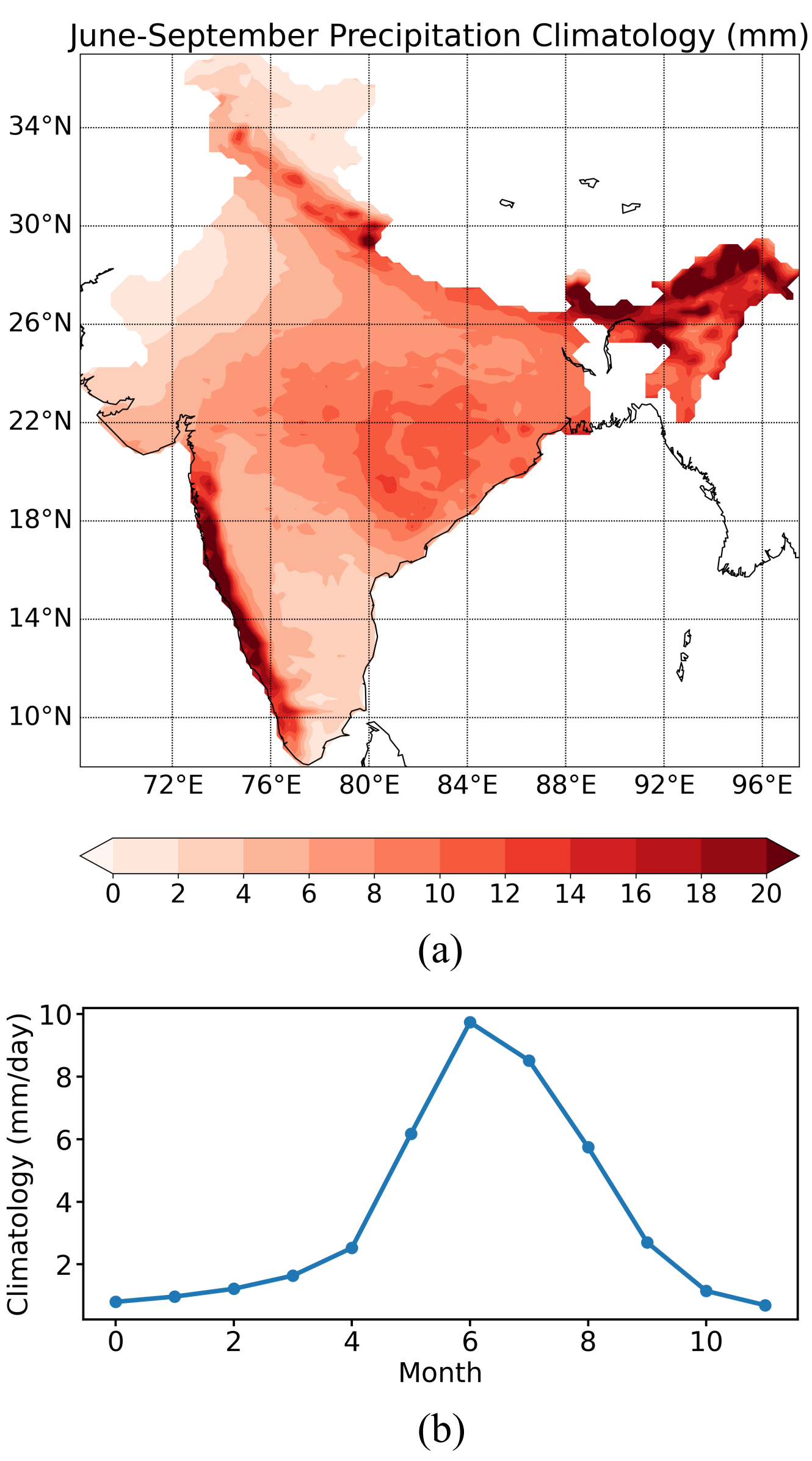

**Figure S3: (a)** Climatological mean June-September rainfall over the Indian monsoon region from ERA5. **(b)** Monthly climatological annual cycle of ISMR (mm/day).

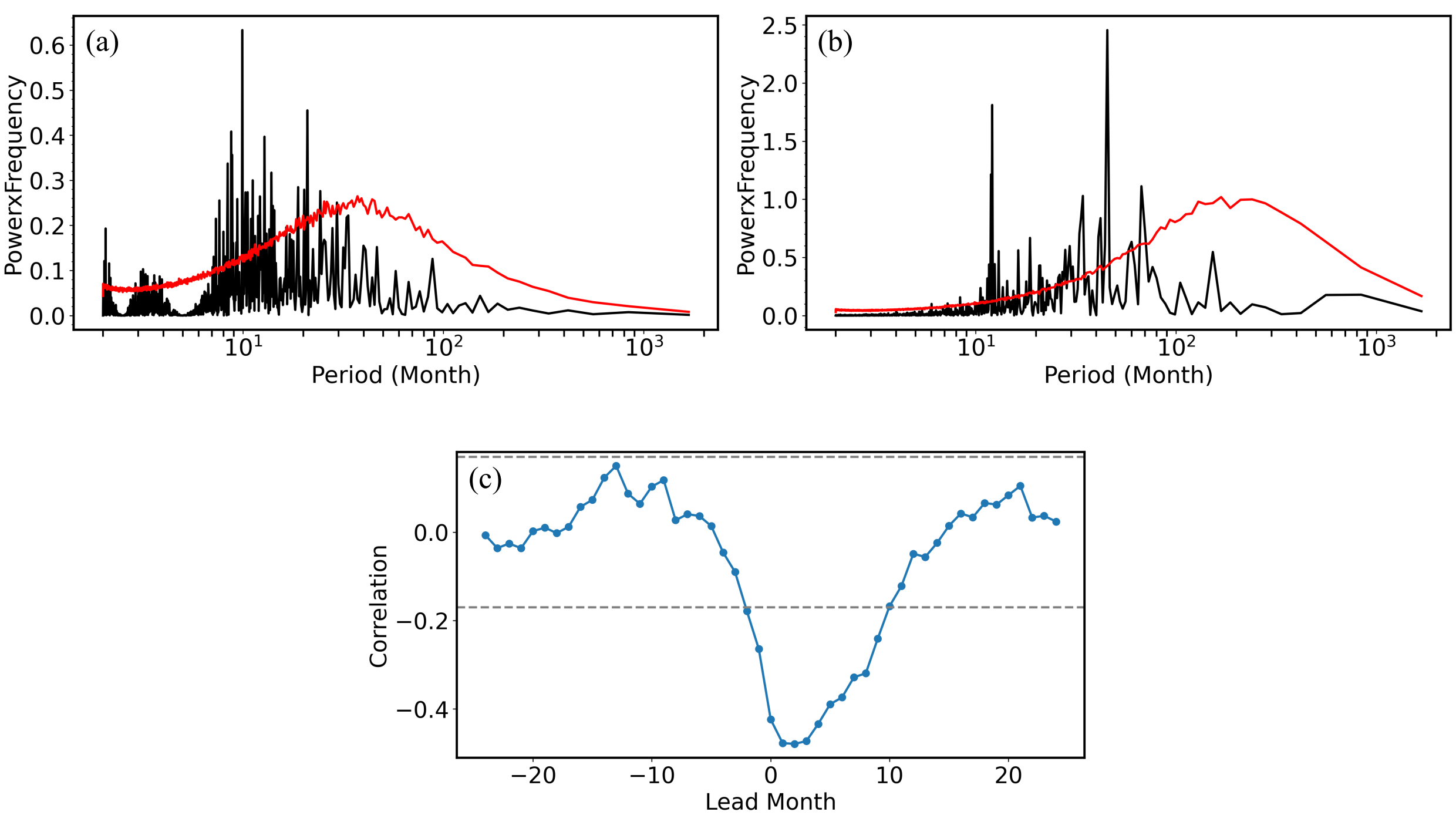

**Figure S4: (a)** Power spectrum of 5-month running mean of monthly rainfall anomalies shown in Figure 1c. **(b)** Power spectrum of first principal component of monthly D20 anomalies shown in Figure 1d. Red noise spectra at 95% confidence level is shown in red. **(c)** Lead-lag correlation between Nino 3.4 index and JJA mean ISMR. COBE (Hirahara et al., 2014) and IITM dataset are used for SST and ISMR, respectively. JJA mean ISMR is obtained between 1873 and 2008.